\documentclass[11pt,a4paper]{article}

\usepackage{jheppub}
\usepackage{epsfig}

\usepackage{graphicx}
\usepackage{color}
\usepackage{latexsym}
\usepackage{amssymb}
\usepackage{slashed}
\usepackage{dcolumn}
\usepackage{axodraw}
\newcommand{\hs}{\hspace*{0.5cm}}
\newcommand{\vs}{\vspace*{0.5cm}}
\newcommand{\be}{\begin{equation}}
\newcommand{\ee}{\end{equation}}
\newcommand{\bea}{\begin{eqnarray}}
\newcommand{\eea}{\end{eqnarray}}

\newcommand{\crn}{\nonumber \\}

\newcommand{\+}{\dagger}
\newcommand{\al}{\alpha}
\newcommand{\la}{\lambda}
\newcommand{\bet}{\beta}

\newcommand{\om}{\omega}

\newcommand{\fr}{\frac}
\newcommand{\bc}{\begin{center}}
\newcommand{\ec}{\end{center}}
\newcommand{\Ga}{\Gamma}

\newcommand{\ep}{\epsilon}

\newcommand{\ta}{\tau}

\newcommand {\ba}{\begin{array}}
\newcommand {\ea}{\end{array}}
\newcommand{\ben}{\begin{enumerate}}
\newcommand{\een}{\end{enumerate}}

\title{ \bf
Probing Dark Matter in the Economical 3-3-1 Model
}

\author[a]{D.~T.~Huong}
\author[b]{C.~S.~Kim}
\author[a]{H.~N.~Long}
\author[b]{and N.~T.~Thuy}

\affiliation[a]{Institute of Physics, VAST, P.O. Box 429, Bo Ho, Hanoi
10000, Vietnam}
\affiliation[b]{Department
  of Physics and IPAP, Yonsei University, Seoul 120-479, Korea}

\emailAdd{dthuong@iop.vast.ac.vn}
\emailAdd{cskim@yonsei.ac.kr}
\emailAdd{hnlong@iop.vast.ac.vn}
\emailAdd{ntthuy@iop.vast.ac.vn}

 \abstract{
We show that the economical  3-3-1 model has a dark mater
candidate. It is a real scalar $H_1^0$ in which main part is
bilepton  (with lepton number 2) and its  mass is in the range of
some TeVs. We calculate the relic abundance of $H_1^0$ dark matter
by using micrOMEGAs 2.4 and figure out parameter space satisfying
the WMAP constraints. Direct and indirect searches are
studied for a special choice of parameters in the WMAP -
allowed region.
 }


 \arxivnumber{1110.1482}

\begin{document}
 \maketitle
 \flushbottom

 \section{Introduction} \setcounter{equation}{0}
\label{sec1}

Now is a golden age of cosmology and astrophysics. Many
abstractive notions such as black holes,
dark matter (DM), dark energy etc. have
become step by step more scientifically
 feasible and widely  accepted subjects.
According to the WMAP~\cite{wmap} the non baryonic dark matter, which is
called the cold dark matter (CDM), must exist
and contain approximately $22\%$ of all energy density of the universe.
The characteristic of the CDM  is thought to be massive and
rarely interact with ordinary matter.
There is yet little astrophysical data
which bear on the CDM.  However, there are a few proposals to
explain the DM in the context of particle physics
~\cite{DM}. The most popular particles
in this class are the sterile neutrino, the axion,
the lightest supersymmetric particle and $etc.$

The DM  does  not exist within the SM.
It has to be realized beyond the SM  at
the electroweak scale or above, so that newly introduced particles in those models
are potentially
good candidates for the DM.
In most supersymmetric models, there is
a conserved multiplicative quantum
number R- parity, which implies that
the lightest superpartner
is stable and can be a DM candidate. This kind of model can not only
explain the original DM but also
represent the
greatest expectations in particle physics  at TeV
scale to be probed by the LHC. However, there is yet no experimental
evidence to support the models. The other way to extend the SM
is the enlargement of the gauge symmetry group from SU(3)$_C\otimes$
SU(2)$_L\otimes$ U(1)$_Y$ to larger groups. In particular, there
exists a simple extension of the SM gauge group to SU(3)$_C\otimes$
SU(3)$_L\otimes$ U(1)$_X$, the so called 3-3-1
models~\cite{331m,331r}.

 Depending on the electric charge of particle at the bottom
of the lepton triplet, the 3-3-1 models are classified into two
main versions: the minimal model \cite{331m} with the lepton
triplet $\left(\nu, l, l^c\right)_L$ and the version with
right-handed (RH) neutrinos  \cite{331r}, where the RH neutrinos
place at the bottom of the triplet: $\left(\nu, l,
\nu^c\right)_L$.  In the 3-3-1 model with right-handed neutrinos,
the scalar sector requires three Higgs triplets. It is interesting
to note that two Higgs triplets of this model have the same
$\mathrm{U}(1)_X$ charges with two neutral components at their top
and bottom. In the model under consideration, the new charge $X$
is connected with the electric charge operator through a relation
\be Q=T_3-\fr{1}{\sqrt{3}}T_8+X.\ee Assigning these neutral
component vacuum expectation values (VEVs) we can reduce the
number of Higgs triplets to  two. Therefore, we have a resulting
3-3-1 model with two Higgs triplets \cite{ecn331w,ecn331}. As a
consequence, the dynamical symmetry breaking also affects the
lepton number. Hence it follows that the lepton number is also
broken spontaneously at a high scale of energy. Note that the
mentioned model contains a very important advantage, namely, there
is no new parameter, but it contains very simple Higgs sector;
therefore, the significant number of free parameters is reduced.
To mark the minimal content of the Higgs sector, this version that
includes right-handed neutrinos (RH$\nu$) is going to be
called the {\it economical 3-3-1 model}.

We would like here to emphasize that by choosing
different electric charge operators, we
can get  a few different versions of 3-3-1
model such as the minimal
3-3-1 model, the 3-3-1 model
with right handed neutrino, the
economical 3-3-1 model and $etc.$ However, all those models have the same
motivations such as
\begin{enumerate}
\item The family number must be a multiplicative  of three.
\item It could explain why the value $\sin^{2} \theta_{W}<\frac{1}{4}$ is observed.
\item It can solve the strong CP problem.
\item It is the simplest model
that includes bileptons of both types: scalar and vectors ones.
\item The model has several sources of CP violation.
\end{enumerate}
Besides those motivations, the 3-3-1
models can contain a candidate for the dark matter.
As an example, the 3-3-1 model with right handed neutrino exhibites that the
scalar bilepton is a candidate for the dark matter as shown in Ref. \cite{biDM}, in which
the authors provided a new quantum number to forbid
the interaction of bilepton with the SM  particles. However, in the
economical 3-3-1 model, which we are considering now,  the model naturally contains a
candidate for the dark matter even without introducing a new quantum
number or a discrete symmetry.

 The paper is organized as follows: Section
\ref{sec2} is devoted to a brief review of the model. The Higgs
sector is considered in the effective approximation $w\gg v,u$ in
section \ref{sec3}. Here, we also find out stable Higgs - a
candidate for the dark matter. The relic abundance as well as its
dependence/indepedence of parameters is figured out in section
\ref{sec4}. We study direct and indirect searches for dark matter
in section \ref{sec5}. Conclusions are given in the last one -
section \ref{sec6}.

\section{A brief review of the  model}
\label{sec2}

The particle content in this model, which is anomaly
free, is given as follows:
\bea \psi_{aL} &=& \left(
               \nu_{aL}, l_{aL}, (\nu_{aR})^c
\right)^T \sim (3, -1/3),\hs l_{aR}\sim (1, -1),\hs a = 1, 2, 3,
\crn
 Q_{1L}&=&\left( u_{1L},  d_{1L}, U_L \right)^T\sim
 \left(3,1/3\right),\crn  Q_{\al L} &= &\left(
  d_{\al L},  -u_{\al L},  D_{\al L}
\right)^T\sim (3^*,0),\hs D_{\al R} \sim \left(1,-1/3\right),\hs
\al=2,3,\crn u_{a R}&\sim&\left(1,2/3\right),\hs d_{a R} \sim
\left(1,-1/3\right),\hs U_{R}\sim \left(1,2/3\right),\eea
where
the values in the parentheses denote quantum numbers based on the
$\left(\mbox{SU}(3)_L,\mbox{U}(1)_X\right)$ symmetry. Unlike the
usual 3-3-1 model with right-handed neutrinos, where the third
family of quarks should be discriminating \cite{longvan}, in this
model under consideration the {\it first} family has to be
different from the two others \cite{dhhl}.
The electric charges of the exotic
quarks $U$ and $D_\al$ are the same as of the usual quarks, $i.e.$,
$q_{U}=2/3$, $q_{D_\al}=-1/3$.

 The spontaneous symmetry breaking in this model is obtained by two
stages: \be \mathrm{SU}(3)_L\otimes \mathrm{U}(1)_X \rightarrow
\mathrm{SU}(2)_L\otimes\mathrm{U}(1)_Y \rightarrow
\mathrm{U}(1)_Q.\ee The first stage is achieved by a Higgs scalar
triplet with a VEV given by \bea \chi=\left(\chi^0_1, \chi^-_2,
\chi^0_3 \right)^T \sim \left(3,-1/3\right),\hs
\langle\chi\rangle=\fr{1}{\sqrt{2}}\left(u, 0, \om
\right)^T.\label{vevc}\eea The last stage is achieved by another
Higgs scalar triplet needed with the VEV as follows \bea
\phi=\left(\phi^+_1, \phi^0_2, \phi^+_3\right)^T \sim
\left(3,2/3\right),\hs
\langle\phi\rangle=\fr{1}{\sqrt{2}}\left(0,v,0
\right)^T.\label{vevp}\eea

 The Yukawa interactions which induce masses for the fermions can
be written in the most general form: \be {\mathcal
L}_{\mathrm{Y}}={\mathcal L}_{\mathrm{LNC}} +{\mathcal
L}_{\mathrm{LNV}},\label{lagrangian} \ee  in which, each part
is defined by \bea
{\mathcal L}_{\mathrm{LNC}}&=&h^U\bar{Q}_{1L}\chi
U_{R}+h^D_{\al\beta}\bar{Q}_{\al L}\chi^* D_{\beta R}\crn
&&+h^l_{ab}\bar{\psi}_{aL}\phi
l_{bR}+h^\nu_{ab}\ep_{pmn}(\bar{\psi}^c_{aL})_p(\psi_{bL})_m(\phi)_n
\crn && +h^d_{a}\bar{Q}_{1 L}\phi d_{a R}+h^u_{\al a}\bar{Q}_{\al
L}\phi^* u_{aR}+ H.c.,\label{y1}\\ {\mathcal
L}_{\mathrm{LNV}}&=&s^u_{a}\bar{Q}_{1L}\chi u_{aR}+s^d_{\al
a}\bar{Q}_{\al L}\chi^* d_{a R}\crn && +s^D_{ \al}\bar{Q}_{1L}\phi
D_{\al R}+s^U_{\al }\bar{Q}_{\al L}\phi^* U_{R}+
H.c.,\label{y2}\eea where the subscripts $p$, $m$ and $n$ stand for
$\mathrm{SU}(3)_L$ indices.

 The VEV $\om$ gives mass for the exotic quarks $U$, $D_\al$ and
the new gauge bosons $Z^{\prime},\ X,\ Y$, while the VEVs $u$ and
$v$ give mass for the quarks $u_a,\ d_a$, the leptons $l_a$ and
all the ordinary gauge bosons $Z,\ W$ \cite{dhhl}.
To keep a consistency with the effective
theory, the VEVs in this model have to satisfy the constraint \be
u^2 \ll v^2 \ll \om^2. \label{vevcons} \ee
 In addition we can derive $v\approx
v_{\mathrm{weak}}=246\ \mbox{GeV}$ and $|u| \leq2.46\ \mbox{GeV}$
from the mass of $W$ boson and the $\rho$ parameter
\cite{ecn331}, respectively. From atomic parity violation in
cesium, the bound for the mass of new natural gauge boson is given
by $M_{Z^{\prime}}>564 \ \mbox{GeV}$ $(\om
> 1400\ \mbox{GeV})$ \cite{dln}. From the analysis on quark
masses, higher values for $\om$ can be required, for example, up
to $10\ \mbox{TeV}$ \cite{dhhl}.

 The Yukawa couplings of (\ref{y1}) possess an extra global
symmetry \cite{changlong,tujo} which is not broken by $ v, \omega$
but by $u$. From these couplings, one can find the following
lepton symmetry $L$ as in Table \ref{lnumber} (only the fields
with nonzero $L$ are listed; all other fields have vanishing $L$).
Here $L$ is broken by $u$ which is behind $L(\chi^0_1)=2$, $i.e.$,
$u$ {\it is a kind of the SLB scale}.

\begin{table}
\bc \caption{\label{lnumber} Nonzero lepton number $L$
 of the model particles.}
\vs
\begin{tabular}{|c|c|c|c|c|c|c|c|c|}

  \hline
   Field
&$\nu_{aL}$&$l_{aL,R}$&$\nu^c_{aR}$ & $\chi^0_1$&$\chi^-_2$ &
$\phi^+_3$ & $U_{L,R}$ & $D_{\alpha L,R}$\\ \hline
  $L$ & $1$ & $1$ & $-1$ & $2$&$2$&$-2$&$-2$&$2$\\
  \hline

\end{tabular}
\ec
\end{table}

It is interesting that the exotic quarks also carry the lepton
number (so-called lepto-quarks); therefore, this $L$ obviously
does not commute with the gauge symmetry. One can then construct a
new conserved charge $\mathcal{L}$ through $L$ by making a linear
combination $L= xT_3 + yT_8 + {\mathcal L} I$. Applying $L$ on a
lepton triplet, the coefficients will be determined
\be L =
\fr{4}{\sqrt{3}}T_8 + {\mathcal L} I \label{lepn}.\ee
Another
useful conserved charge $\mathcal B$, which is exactly not broken
by $u$, $v$ and $\om$, is usual baryon number: $B ={\mathcal B} I$.
Both the charges $\mathcal{L}$ and $\mathcal{B}$ for the fermion
and Higgs multiplets are listed in Table~\ref{bcharge}.

\begin{table}
\bc \caption{\label{bcharge} ${\mathcal B}$ and ${\mathcal L}$ charges
of the model multiplets.}
\vs
\begin{tabular}{|c|c|c|c|c|c|c|c|c|c|c|}
  \hline
  Multiplet & $\chi$ & $\phi$ & $Q_{1L}$ & $Q_{\al L}$ &
$u_{aR}$&$d_{aR}$ &$U_R$ & $D_{\al R}$ & $\psi_{aL}$ & $l_{aR}$
\\ \hline
  $\mathcal B$-charge &$0$ & $ 0  $ &  $\fr 1 3  $ & $\fr 1 3
$& $\fr 1 3  $ &
 $\fr 1 3  $ &  $\fr 1 3  $&  $\fr 1 3  $&
 $0  $& $0$ \\ \hline
  $\mathcal L$-charge &$\fr 4 3$ & $-\fr 2 3  $ &
   $-\fr 2 3  $ & $\fr 2 3  $& 0 & 0 & $-2$& $2$&
 $\fr 1 3  $& $ 1   $ \\
  \hline
\end{tabular}

\ec
\end{table}

Let us note that the Yukawa couplings of (\ref{y2}) conserve
$\mathcal{B}$, however, violate ${\mathcal L}$ with $\pm 2$ units
which implies that these interactions are much smaller than the
first ones \cite{dhhl}: \be s_a^u, \ s_{\al a}^d,\ s_\al^D, \
s_\al^U \ll h^U,\ h_{\al \bet}^D,\ h_a^d,\ h_{\al
a}^u.\label{dkhsyu}\ee

In this model, the most general Higgs potential has very simple
form \cite{dls} \bea V(\chi,\phi) &=& \mu_1^2 \chi^\dag \chi +
\mu_2^2 \phi^\dag \phi + \la_1 ( \chi^\dag \chi)^2 + \la_2 (
\phi^\dag \phi)^2\crn & & + \la_3 ( \chi^\dag \chi)( \phi^\dag
\phi) + \la_4 ( \chi^\dag \phi)( \phi^\dag \chi). \label{poten}
\eea It is noteworthy that $V(\chi,\phi)$ does not contain
trilinear scalar couplings and conserves both the mentioned global
symmetries, this makes the Higgs potential much simpler and
discriminative from the previous ones of the 3-3-1 models
\cite{changlong,tujo,ochoa2}. The non-zero values of $\chi$ and
$\phi$ at the minimum value of $V(\chi,\phi)$ can be obtained
by\bea \chi^\dag\chi&=&\fr{\lambda_3\mu^2_2
-2\lambda_2\mu^2_1}{4\lambda_1\lambda_2-\lambda^2_3}
\equiv\fr{u^2+\om^2}{2},\label{vev1}\\
\phi^\dag\phi&=&\fr{\lambda_3\mu^2_1
-2\lambda_1\mu^2_2}{4\lambda_1\lambda_2-\lambda^2_3}
\equiv\fr{v^2}{2}.\label{vev2}\eea Any other choice of $u,\ \om$
for the vacuum value of $\chi$ satisfying (\ref{vev1}) gives the
same physics because it is related to (\ref{vevc}) by an
$\mbox{SU}(3)_L\otimes \mbox{U}(1)_X$ transformation. It is worth
noting that the assumed $u\neq 0$ is therefore given in a general
case.  This model of course leads to the formation of Majoron
\cite{dls}.

\section{Stable Higgs bosons in 3-3-1 model}
\label{sec3}

This section is to show that the economical $3-3-1$ model
furnishes a good candidate for self interacting dark matter. The
important properties are that dark matter must be stable and
neutral. Hence, we are going to consider the scalar sector of the
model and specially neutral scalar sector, and we can specify
whether any of them can  satisfy the self interacting dark matter
conditions.

 Let us review the Higgs states and coupling constants. In this
model, the most general Higgs potential has very simple form given
in (\ref{poten}).
As usual, we first
shift the Higgs fields as follows:
\be \chi=\left(%
\begin{array}{c}
  \chi^{P 0}_1 + \fr{u}{\sqrt{2}}  \\
  \chi^-_2 \\
  \chi^{P 0}_3 + \fr{\om}{\sqrt{2}}  \\
\end{array}%
\right),  \hs  \phi=\left(%
\begin{array}{c}
  \phi^{+}_1   \\
  \phi^{P 0}_2 + \fr{v}{\sqrt{2}}\\
  \phi^{+}_3   \\
\end{array}%
\right).\label{higgsshipt} \ee The subscript $P$ denotes {\it
physical} fields as in the usual treatment. The
constraint equations at the tree level are given as \bea \mu_1^2 + \la_1 (u^2 +
\om^2) + \la_3 \fr{v^2}{2}
 & = & 0,\label{potn1}\\
 \mu_2^2 +  \la_2 v^2   + \la_3 \fr{(u^2 + \om^2)}{2} & = &
0.\label{potenn2} \eea
Note that $u$ is a parameter of lepton-number violation, therefore the
terms linear in $u$ violate the latter.  Applying the constraint
equations (\ref{potn1}) and (\ref{potenn2}) we get the minimum
value, mass terms, lepton-number conserving and violating
interactions as follows \bea V(\chi,\phi)
&=&V_{\mathrm{min}}+V^{\mathrm{N}}_{\mathrm{mass}}
+V^{\mathrm{C}}_{\mathrm{mass}}+V_{\mathrm{LNC}} + V_{\mathrm{LNV}},
\label{potenn2a}\eea where \bea V_{\mathrm{min}}&=&- \fr{\la_2}{4}
v^4 - \fr 1 4 (u^2+\om^2)[\la_1(u^2 + \om^2) + \la_3 v^2],\crn
V^{\mathrm{N}}_{\mathrm{mass}}&=& \la_1 (uS_1+\om S_3)^2+\la_2 v^2
S^2_2 +\la_3 v (uS_1+\om S_3)S_2,
\label{potenn8}\\
V^{\mathrm{C}}_{\mathrm{mass}}&=&\fr{\la_4}{2}(u\phi^+_1+v\chi^+_2+\om
\phi^+_3)(u\phi^-_1+v\chi^-_2+\om \phi^-_3),
\label{potenn12}\\
V_{\mathrm{LNC}} &= &\la_1
(\chi^\+\chi)^2+\la_2(\phi^\+\phi)^2+\la_3
(\chi^\+\chi)(\phi^\+\phi)+\la_4 (\chi^\+\phi)(\phi^\+\chi)\crn &&
+ 2\la_1\om S_3(\chi^\+\chi)+2\la_2 v S_2(\phi^\+\phi)+\la_3 v
S_2(\chi^\+\chi)+\la_3\om S_3(\phi^\+\phi) \crn
&&+\fr{\la_4}{\sqrt{2}}(v\chi^-_2+\om
\phi^-_3)(\chi^\+\phi)+\fr{\la_4}{\sqrt{2}}(v\chi^+_2+\om
\phi^+_3)(\phi^\+\chi),  \label{potenn3} \\ V_{\mathrm{LNV}} &= &
2 \la_1 u S_1(\chi^\+\chi)+\la_3u
S_1(\phi^\+\phi)+\fr{\la_4}{\sqrt{2}}u
\left[\phi^-_1(\chi^\+\phi)+\phi^+_1(\phi^\+\chi)\right].
\label{potenn4} \eea In the above equations, we have dropped the
subscript $P$ and used $\chi=(\chi^{0}_1,\chi^-_2,\chi^{0}_3)^T$,
$\phi=(\phi^{+}_1,\phi^{0}_2,\phi^{+}_3)^T$. Moreover, we have
expanded the neutral Higgs fields as \be \chi^0_1  =  \fr{S_1 +i
A_1}{\sqrt{2}},\hs \chi^0_3  =  \fr{S_3 + i A_3}{\sqrt{2}},\hs
\phi^0_2 = \fr{S_2 + i A_2}{\sqrt{2}}.\label{potenn5}\ee In the
pseudoscalar sector, all the fields are Goldstone bosons: $G_1=
A_1$, $G_2= A_2$ and $G_3 = A_3$ (cl.  Eq. (\ref{potenn8})). The
scalar fields $S_1$, $S_2$ and $S_3$ gain masses via
(\ref{potenn8}), thus we get one Goldstone boson $G_4$ and two
neutral physical fields the standard model $H^0$ and the new
$H^0_1$ with masses  \bea m^2_{H^0}&=&\la_2 v^2+
\la_1(u^2+\om^2)-\sqrt{[\la_2 v^2-\la_1(u^2+\om^2)]^2+\la^2_3 v^2
(u^2+\om^2)}\crn
&\simeq&\fr{4\la_1\la_2-\la^2_3}{2\la_1}v^2,\label{potenn10a}\eea
\bea M^2_{H^0_1}&=&\la_2 v^2+ \la_1(u^2+\om^2)+\sqrt{[\la_2
v^2-\la_1(u^2+\om^2)]^2+\la^2_3 v^2 (u^2+\om^2)}\crn &\simeq&
2\la_1 \om^2.\label{potenn10}\eea  In term of original fields, the
Goldstone and Higgs fields are given by \bea G_4 &=&
\fr{1}{\sqrt{1+t^2_{\theta}}}(S_1 -t_\theta
S_3),\\
H^0&=&c_\zeta S_2
-\fr{s_\zeta}{\sqrt{1+t^2_{\theta}}}(t_{\theta}S_1 +S_3),
\label{potenn11a}\\
H^0_1&=&s_\zeta S_2
+\fr{c_\zeta}{\sqrt{1+t^2_{\theta}}}(t_{\theta}S_1
+S_3),\label{potenn11b}\eea where \bea t_{2\zeta}&\equiv &
\fr{\la_3 M_W M_X}{\la_1M^2_X-\la_2 M^2_W}.\label{potenn11}\eea
{}From Eq. (\ref{potenn10}), it follows that mass of the new Higgs
boson $M_{H^0_1}$ is related to mass of the bilepton gauge $X^0$
(or $Y^\pm$ via the law of Pythagoras) through \bea M^2_{H^0_1}&
=& \fr{8\la_1}{g^2} M_X^2 \left[1 +
\mathcal{O}\left(\fr{M_W^2}{M_X^2}\right)\right]\crn &=& \fr{2
\la_1 s^2_W}{\pi \al} M_X^2  \left[1 +
\mathcal{O}\left(\fr{M_W^2}{M_X^2}\right)\right] \approx 18.8
\la_1 M_X^2. \label{potenn11mass}\eea Here, we have used $\al =
\fr{1}{128}$ and $s^2_W = 0.231$.

In the charged Higgs sector, the mass terms for
$(\phi_1,\chi_2,\phi_3)$ are given by (\ref{potenn12}), thus there
are two Goldstone bosons and one physical scalar field:\be
H^+_2\equiv \fr{1}{\sqrt{u^2+v^2+\om^2}}(u\phi^+_1+v\chi^+_2+\om
\phi^+_3)\label{potenn13}\ee with mass \bea M^2_{H^+_2}&
=&\fr{\la_4}{2}(u^2+v^2+\om^2) = 2 \la_4 \fr{M_Y^2}{g^2} = \fr{s_W^2
\la_4}{2 \pi \al} M_Y^2 \simeq 4.7 \la_4 M_Y^2.\label{potenn14}\eea
The two remaining  Goldstone bosons are \bea
G^+_5&=&\fr{1}{\sqrt{1+t^2_{\theta}}}(\phi^+_1-t_{\theta}\phi^+_3),\\
G^+_6
&=&\fr{1}{\sqrt{(1+t^2_{\theta})(u^2+v^2+\om^2)}}\left[v(t_\theta
\phi^+_1+\phi^+_3)- \om
(1+t^2_\theta)\chi^+_2\right].\label{potenn15}\eea

Thus, all the pseudoscalars are eigenstates and massless
(Goldstone). Other fields are related to the scalars in the weak
basis by the linear transformations: \be \left(
\begin{array}{ccc} H^0
\\ H^0_1 \\G_4
\end{array}\right) = \left( \begin{array}{ccc} -s_\zeta
s_\theta & c_\zeta
&  -s_\zeta c_\theta\\
c_\zeta s_\theta &
 s_\zeta & c_\zeta c_\theta\\
c_\theta & 0 & -s_\theta \end{array}\right) \left(
\begin{array}{ccc} S_1
\\ S_2  \\ S_3 \end{array}\right), \label{potenn16}\ee \be
 \left(%
\begin{array}{c}
  H^+_2 \\
  G^+_5 \\
  G^+_6
\end{array}%
\right)=\fr{1}{\sqrt{\om^2+c^2_\theta v^2}}\left(%
\begin{array}{ccc}
  \om s_\theta & vc_\theta & \om c_\theta\\
  c_\theta \sqrt{\om^2+c^2_\theta v^2} & 0 & -
  s_\theta \sqrt{\om^2+c^2_\theta v^2}\\
  \fr{v s_{2\theta}}{2} & -\om & v c^2_{\theta}
\end{array}%
\right)\left(%
\begin{array}{c}
  \phi^+_1 \\
  \chi^+_2 \\
  \phi^+_3
\end{array}%
\right).\label{potenn18}\ee
Let us comment on our physical Higgs
bosons. In the effective approximation $w \gg v, u$, from Eqs.
(\ref{potenn16}),
 and (\ref{potenn18}) it follows that
 \bea H^0 &\sim &
S_2,\hs H_1^0 \sim S_3, \hs G_4 \sim S_1, \crn H^+_2 &\sim &
\phi_3^+,\hs G^+_5 \sim \phi^+_1, \hs G^+_6 \sim
\chi^+_2.\label{potenn20a} \eea
{}From the  Higgs gauge interactions
given in \cite{dls}, the coupling constants of $
H_1^0 $ Higgs and SM gauge bosons depend on $s_\zeta$ with
$t_{2\zeta} =\fr{\lambda_3 M_W M_X} {\lambda_1 M_X^2 -\lambda_2
M_W^2} $. In the  $w \gg v, u$ limit, $M_X \gg M_W$ or
$|t_{2\zeta}|\rightarrow 0$. Therefore, the $H_1^0$ Higgs does not
interact with the SM gauge bosons $W^\pm, Z^0, \gamma$. However,
there are couplings of $H_1^0$ Higgs with the Bilepton $Y$ and
$Z^\prime$. In order to forbid the decay of $H_1^o$, we assume that
$M^2_{H_1^0} \leq M^2_Y$. It means that $2 \lambda_1 \omega^2 \leq
\frac{1}{4} g^2\omega^2 $ or $\lambda_1 \leq 0.051$.
The interactions of $ H_1^0$ Higgs with new gauge boson $Z^\prime
$ is $Z^\prime-H_1^0- G_3$ interaction. But $G_3$ is a Goldstone
bosons, this interaction can be gauged away by a unitary
transformation.

 Let us consider the interaction of dark matter to Higgs bosons.
{}From the Higgs potential (\ref{poten}), we can obtain the
coupling of the new Higgs $H^0_1$  to $H^0 H^0 $. The decay
rate of the $H_1^0 \rightarrow H^0 H^0$
is written as
\bea \Ga _{H_1^0 \rightarrow H^0
H^0}=\fr{\lambda_3^2}{16\pi} \fr{w^2}{M_{H_1^0}}\left(
1-\fr{2M_{H^0}^2}{M_{H_1^0}^2}\right). \eea
The lifetime is the inversion  of decay rate
 $\ta =\fr{\hbar}{\Ga}$, with $ \hbar=6.6 \times 10^{-25}\textrm{ GeV}
 \times s $.
  If taking $\ta > 10^{20}s$ (the life time longer than our universe's age),
  $M_{H_1^0}= 7000\, \textrm{GeV},
 w= 10 000\, \textrm{GeV}, M_{H^0}= 120\, \textrm{GeV} $, then we get
\be \Ga _{H_1^0 \rightarrow H^0 H^0}= \fr{\lambda_3^2}{16\pi}
\fr{10^8}{7.10^3}\left( 1-\fr{2 \times
120^2}{49.10^6}\right)\simeq 284 \times \lambda_3^2. \ee
In order to get the constraint on the lifetime of $H_1^0$
longer than our universe's age, it is easy to see that the value of
$\lambda_3$ is approximately order of $ 10^{-24}$. It is to be
emphasized that the limit of $\lambda_3$ makes sure that $t_\zeta$
is small.

To avoid $H_1^0$ decaying to $H_2^+ H_2^-$, we need the constraint
for the mass of two Higgs, namely $M^2_{H_1^0}<4M^2_{H_2^+}$, which
means $\lambda_1 < \lambda_4$.
{}From the Lagrangian given in
(\ref{lagrangian}), it is easy to see  that the $H_1^0$ does not  interact
with the SM leptons but it interacts with exotic quarks. As we
know the exotic quarks are heavy ones, we assume that their masses
are heavier than that of $H_1^0$. Hence, $H_1^0$ can be stable and
be candidate for dark matter.

\section{Thermal relic abundance}
\label{sec4}

\subsection{Constraints}
\label{constraints}

Before considering the relic abundance of dark matter, let us
summarize the constraints on the couplings $\la_{1,2,3,4}$, the
VEV $ w$, and exotic quarks masses:
 \ben
\item  From  Eqs. (\ref{potenn10a}) and (\ref{potenn10}),
 we obtain the constraints as follows
 \be \la_1 > 0, \ \la_2 > 0, \hs 4 \la_1 \la_2 > \la_3^2. \label{potenn20}\ee
\item  The mass of the charged Higgs boson
$H^\pm_2$ is proportional to that of the charged bilepton $Y$
through a coefficient of Higgs self-interaction $\la_4>0$.
Analogously, this happens for the standard-model-like Higgs boson
$H^0$  and the new $H^0_1$ . Combining (\ref{potenn20})
with the constraint equations
(\ref{potn1}), (\ref{potenn2}) we get a consequence: $\la_3$ is
negative ($\la_3 < 0$).
\item In order to get the stable Higgs particle $H_1^0$,
we need the constraints as follows\\
\be \la_1 < \la_4, \hs  \la_1 \leq 0.051,\hs  |\lambda_3 |\sim
10^{-24},\hs  M_{H_1^0}\leq M_U. \label{hu1}\ee

Since $\la_3 < 0$, we get $\lambda_3 \sim -10^{-24}$.

\item In the limit of $\lambda$ given in (\ref{hu1}),
the SM  Higgs mass can be estimated as $M_{H^0}^2= \frac{4
\lambda_1 \lambda_2 -\lambda_3^2}{2\lambda_1}v^2 \simeq 2\lambda_2
v^2 $. Combining with the constraint 80 $< M_{H^0} <$160 GeV, we can
obtain the constraints on $\lambda_2$  as follows: $0.053<
\lambda_2 < 0.212$. \een

\subsection{Implication for parameter space from the WMAP constraints.}

In this subsection, we discuss constraints on the parameter space of
the 3-3-1 model originating from the WMAP results on dark matter
relic density \cite{WMAPconstraint},
$$ \Omega h^2 = 0.1120 \pm 0.0056 ~.$$
In order to calculate the
relic density, we use micrOMEGAs 2.4 \cite{micrOmegas2.4} after
implementing new model files into CalcHEP \cite{CalcHEP}. The
parameters of our model are the self-Higgs couplings, $\lambda_1,
 \lambda_2, \lambda_3, \lambda_4,$ the VEV $w$ and exotic quarks
masses. Note that the usual quarks $u, s, b$ gain masses at
one-loop level \cite{dhhl}. We express the couplings of
Higgs with exotic quarks $s^u_1, s^d_{22}, s^d_{33} $ as functions
of $\lambda_1,~ w$ and exotic quarks masses.

All Feynman diagrams contributing to
the annihilation of $H_1^0$ Higgs are listed in the Appendix (\ref{appendix1}).
At the tree level, the annihilation of $H_1^0$ dark matter can be
done through s-, or t-channel, or direct annihilation. Since there
is neither coupling of $H_1^0 H_1^0$ to one neural gauge boson nor
coupling of $H_1^0 H_1^0$ with one fermion, the propagator in the
s-channel can be $H_1^0$ or $H^0$ Higgs only. To draw Feynman
diagrams contributing to the annihilation of $H_1^0 H_1^0$ through
s- and t-channels, we list all non-zero couplings $H_1^0 AB $ and
$H^0 AB $, where $A,B$ can be Higgs, or gauge boson, or fermion.
We see that $H_1^0$ couples to one usual quark (anti-usual quark)
$u, s, b$ and one anti-exotic quark (exotic quark) while  $H^0$
couples to $c, d, t$ quarks. Therefore, the annihilation of $H_1^0
H_1^0$ into $u\bar u$, $s\bar s$, $b\bar b$ are done through t-channel
through  exotic quark exchange while the contributions of
the remaining usual quarks are done via s-channel through $H^0$ exchange.
Since the coupling $H_1^0 H_1^0H^0 \sim v \la_3$, the contribution of
$c \bar c$, $d\bar d$, $t\bar t$ channels to $\fr {1}{\Omega h^2}$ is very small.
$H_1^0$ Higgs  can also annihilate into two Higgs bosons or two gauge
bosons directly. Theses vertices arise from the Higgs potential
and Higgs-gauge interactions.

\begin{figure}[h]
\begin{center}
\includegraphics[width=7.5cm,height=7cm] {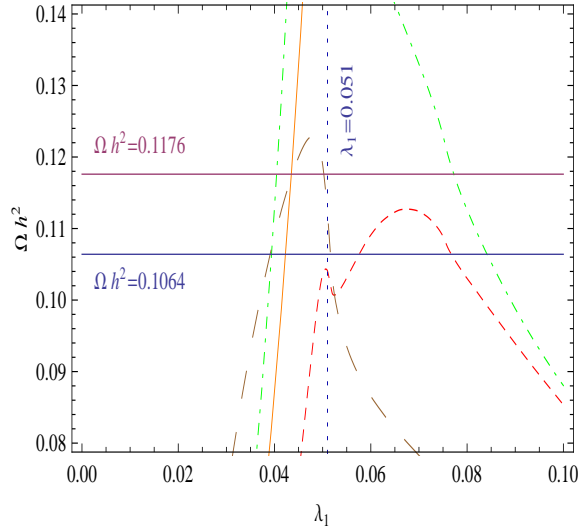}
\caption{\label{vary-la1}{$\Omega h^2 $ vs $\la_1$ for
$\la_2$ = 0.12, $\la_3$ = $-10^{-24}$, $\la_4$ = 0.06,
 $M_{D_2}$ =$M_{D_3}$=100 TeV, and for $w$=10 TeV, $M_U$ = 24 TeV
 (red dashed line),  $w$ = 10 TeV, $M_U$ = 36 TeV
 (green dot-dashed line),
 $w$ = 30 TeV,  $M_U$ = 36 TeV (orange full line),
 and  $M_{D_2}$ = $M_{D_3}$ = 12 TeV, $w$ = 10 TeV, $M_U$ = 70 TeV
 (brown large dashing line).
 The blue dotted vertical line corresponds to $\la_1$ = 0.051. }}
\end{center}
\end{figure}

First we study the
behavior of $\Omega h^2$ as a function of one parameter each time.
Table \ref{vary-lamda} shows the dependence of $\Omega h^2$
on $\la_2$, $\la_3$, $\la_4$ corresponding to the point 1, 2, 3.
In all three cases, we fix $\la_1=0.04$, $w=10$ TeV,
and exotic quarks masses $M_U=36$ TeV, $M_{D_2}=M_{D_3}=100$ TeV
as a special choice of parameters in the WMAP allowed band
(please look at Fig. \ref{vary-la1}),
the green dot-dashed line. We can see that neither $\Omega h^2$ nor
the contribution of channels change when varying $\la_2$ in
the range 0.053 $\sim$ 0.212,
or $\la_3$ from $-10^{-33}$ to $-10^{-20}$, which regions satisfy
the constraints given in \ref{constraints}.
The couplings $H^0H^0H^0$ and $H^0H_2^+H_2^-$ are proportional to
$v\la_2$, and these contribute to the annihilation of $H_1^0$ dark
matter through s-channel $H$ exchange. The coupling
$H_1^0H_1^0H^0$ is $v\la_3\sim \la_3$, where $\la_3$ is very
small. That is why $\Omega h^2$ does not depend on $\la_2$.
The relic density changes negligibly if we vary $\la_4$.
It gets the  minimum value $\Omega h^2=0.1116$ at
the point $\la_4=0.15$  and the maximum value $\Omega h^2=0.1128$
at the two points $\la_4=0.08$ and $\la_4=0.19$.
With $\la_4 \geq 0.2$, the relic density keeps constant value $\Omega h^2=0.1095$.


\begin{table}
\bc \caption{\label{vary-lamda}$\Omega h^2$ and dominant channels
when varying $\la_2,\la_3, \la_4 $.  }
\vs
\begin{tabular}{|c|c|c|c|}
  \hline
  point & 1 & 2 & 3 \\ \hline
  $\la_1$& 0.04 & 0.04 & 0.04  \\ \hline
  $\la_2$ & 0.053 to 0.212 & 0.12 & 0.12 \\ \hline
  $\la_3$ & $-10^{-24}$ & $-10^{-33}$ to $-10^{-20}$& $-10^{-24}$ \\ \hline
  $\la_4$ & 0.06 & 0.06 & 0.004 to 200 \\ \hline
 $w$ (TeV)& 10  & 10  & 10  \\ \hline
   $M_U$ (TeV)& 36 & 36  & 36  \\ \hline
  $M_{D_2}$(TeV) & 100 & 100  & 100    \\ \hline
  $M_{D_3}$(TeV) & 100  & 100  &100    \\ \hline
  $\Omega h^2$  & 0.1127 & 0.1127 & (0.1116, 0.1128) \\ \hline
          & $u\bar u (97.40\%)$ & $u\bar u (97.40\%)$ & $u\bar u (97.36-97.45\%)$ \\
          & $s\bar s (1.26\%)$ &  $s\bar s (1.26\%)$ &  $s\bar s (1.26\%)$   \\
channels  & $b \bar b (1.28\%)$  & $b \bar b (1.28\%)$ & $b \bar b (1.28\%)$ \\
          & $H_2^+ H_2^- (0.05\%)$  & $H_2^+ H_2^- (0.05\%)$  & $H_2^+ H_2^- (0-0.09\%)$\\
          & rest $(0.01\%)$ &  rest $(0.01\%)$ & rest $(0.01-0.02\%)$
  \\ \hline
\end{tabular}
\ec
\end{table}

 Now we consider the dependence of the relic density of $H_1^0$
dark matter on the remaining parameters $\la_1, w, M_U,
 M_{D_2}$ and $  M_{D_3}$.
First, we fix the values of $\lambda_{2,3,4}$ satisfying  the
constraints given in (\ref{hu1}), especially taking $\la_2$ =
0.12, $\la_3 = -10^{-24}, \la_4$ = 0.06 and varying the masses of
exotic quarks. We consider the relic density as a
 function of $\la_1$. Fig. \ref{vary-la1} compares WMAP data to
 the theoretical prediction. The red dashed line presents predictions from our theory
by fixing $M_{D_2}$ = $M_{D_3}$ = 100 TeV, $w$=10 TeV and $M_U$ = 24
 TeV. In order to meet fully
 the WMAP dada, the value of $\lambda_1$ must be
 different from  the allowed value in (\ref{hu1}). However, if we
 change the masses of exotic quarks, we can obtain allowed region,
 namely the green dot-dashed line given by taking
 $w= 10$ TeV  and $M_U = 36$ TeV. The allowed region of $\lambda_1$
 satisfy both  the WMAP data and the stable Higgs constraints
 (\ref{hu1}) is 0.0393 $ < \lambda_1 <$ 0.0406.
 The orange full line is obtained by fixing $w=30$ TeV
 and $M_U =36$ TeV. In this case, the constraints on $\lambda_1$
 is 0.0424  $ < \lambda_1 <$  0.0436. On the other hand, if we change the mass
 of exotic $D$-quarks, we can find the other allowed region of
 $\lambda_1$. For example, if we take
$M_{D_2}$ = $M_{D_3}$ = 12 TeV, the allowed region of $\lambda_1$
is 0.0502  $ < \lambda_1 <$ 0.051. Hence, we could conclude that the mass of exotic
$U$-quark can be larger or smaller than that of $D$-quarks in order
to come to agreement with the WMAP data.

\begin{figure}[h]
\begin{center}
\includegraphics[width=7.5cm,height=7cm]{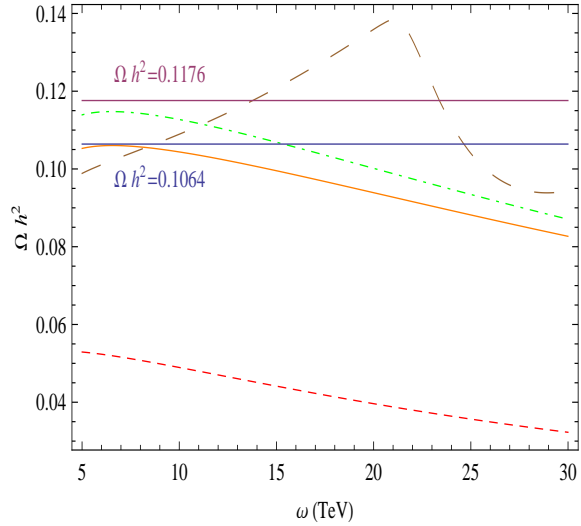}
\caption{\label{vary-w}{$\Omega h^2 $ vs $w$ for
$\la_1$ = 0.04, $\la_2$ = 0.12, $\la_3$ = $-10^{-24}$, $\la_4$ = 0.06, and
 for $M_U$ = 24 TeV, $M_{D_2}$ = $M_{D_3}$ = 100 TeV (red dashed line),
 $M_U$ = 36 TeV, $M_{D_2}$ = $M_{D_3}$ = 100 TeV (green dot-dashed line),
 $M_U = M_{D_2} = M_{D_3}$ = 36 TeV (orange full line) and
 $M_U$ = 70 TeV, $M_{D_2} = M_{D_3}$ = 12 TeV (brown large dashing line).}}
\end{center}
\end{figure}

Fig. \ref{vary-w} shows the dependence of the relic density on the
VEV $w$ for $\la_1$ = 0.04, $\la_2$ = 0.12, $\la_3$ = $-10^{-24}$
and $\la_4$ = 0.06.
 This figure shows that the VEV $w< $ 15.33 TeV is in the
 WMAP-allowed region for $M_U$ = 36 TeV, $M_{D_2}$= $M_{D_3}$
 =100 TeV. However, if the values of $M_U=24$ TeV or $M_D=M_U=36$ TeV, there is no
 allowed region of $\omega$ in agreement with the WMAP data.
 The situation becomes  totally different for $M_U$ = 70 TeV,
 $M_{D_2}$ = $M_{D_3}$ = 12 TeV (brown large dashing line).
 The relic density at first increases
 then decreases as a function of $w$. In
 the WMAP band, $w$ is in the range 8.752 - 13.85 GeV or 23.3 - 24.61 GeV.

Similarly, we can figure out the region of $M_U$, $M_{D_2}$, and
$M_{D_3}$ in agreement with the WMAP data by fixing $\la_1$ =
0.04, $\la_2$ = 0.12, $\la_3$ = $-10^{-24}$, $\la_4$ = 0.06, and
$w$ = 10 TeV. The value of $M_U$ is in the narrow band 34.93
$\sim$ 36.73 TeV  for $M_{D_2}$ = $M_{D_3}$ = 100 TeV, and $M_U$
is should be heavier, 66.71 $< M_U < $ 85.01 TeV if we take
$M_{D_2}$ = $M_{D_3}$ = 12 TeV. In case of $M_U$ = 36 TeV and
$M_{D_3}$ = 100 TeV, $M_{D_2}$ is in the region of 37.99 $<
M_{D_2} <$ 259.7 TeV. For $M_U$ = 36 TeV and $M_{D_2}$ = 100 TeV,
the relic density is always in the WMAP-allowed region for
$M_{D_3} >$ 40 TeV. There are many choices of ($M_U$, $M_{D_2}$,
$M_{D_3}$) set satisfying the WMAP result, and we can see that the
WMAP constraints do not require the order of exotic
quarks masses.

\begin{figure}[h]
\begin{center}
\includegraphics[width=7.5cm,height=7.5cm]{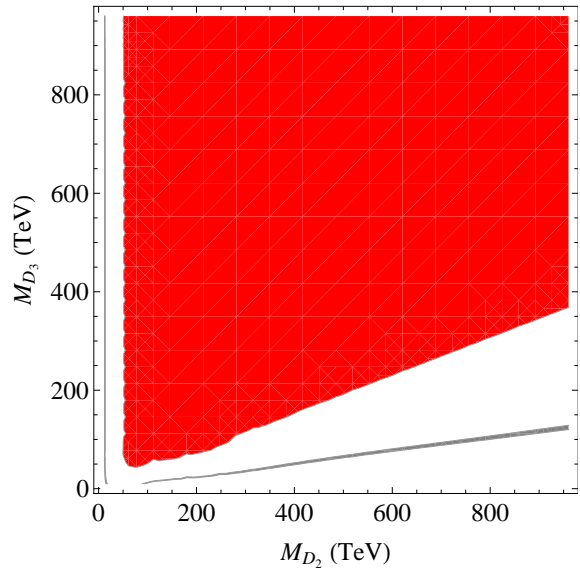}
\caption{\label{vary-MD2-MD3}{Contour plots for $0.1064 < \Omega h^2
< 0.1176 $ (WMAP constraints)
in $M_{D_2}-M_{D_3}$ plane for $\la_1$ = 0.04, $\la_2$ = 0.12,
$\la_3 = -10^{-24}$, $\la_4$ = 0.06, $w$=10 TeV, and for $M_U$ = 36 TeV
 (red) and $M_U$ = 70 TeV (grey).}}
\end{center}
\end{figure}

 To give an overview of the behavior of the relic density
in the $M_{D_2}$ - $M_{D_3}$ plane,  as shown in Fig. \ref{vary-MD2-MD3},
we consider the model with $\la_1$ = 0.04, $\la_2$= 0.12,
$\la_3 = -10^{-24}, \la_4$ = 0.06 and $w$ = 10 TeV.  The region of
$M_{D_2}$ - $M_{D_3}$ in agrement with the WMAP is very wide for
$M_U$ = 36 TeV (red), while it
 seems to be two narrow bands for
$M_U$ = 70 TeV (grey bands).

\begin{figure}[h]
\begin{center}
\includegraphics[width=7cm,height=7cm]{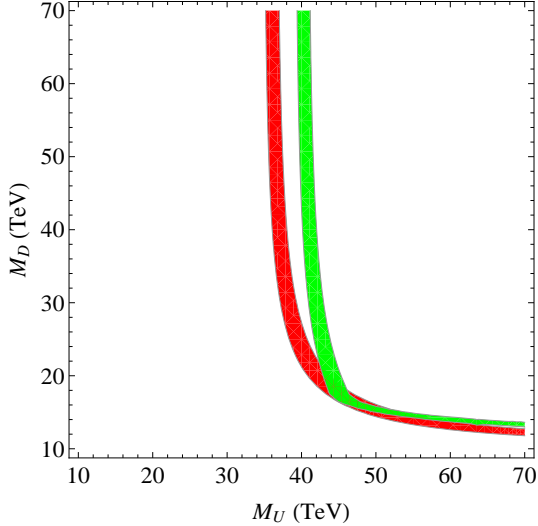}
\caption{\label{vary-MU-MD}{Contour plots for $0.1064 < \Omega h^2
< 0.1176 $ (WMAP constraints)
in  $M_U-M_D$ plane for $\la_1$=0.04, $\la_2$=0.12,
$\la_3 = -10^{-24}$, $\la_4$=0.06 and for $w$=10 TeV (red) and
 $w$=30 TeV (green).}}
\end{center}
\end{figure}

{}From now on we take $M_D$ = $M_{D_2}$ = $M_{D_3}$ for
convenience. In the $M_U$ - $M_D$ plane (see Fig. \ref{vary-MU-MD}), we
can see that to satisfy the WMAP constraints, $M_U$ must be heavier than
$35.2$ TeV and $M_D$ must be heavier than $11.8$ TeV. For $M_U =36$ TeV
and $w= 10$ TeV, the relic density as slowly varying function of
$M_D$. The situation is similar to that of $M_U= 40$ TeV and
$w=30$ TeV. On the other hand, for the value of $M_D$  around 12
TeV, the relic density varies very slowly as a function of $M_U$. For
$12 < M_D < 22$ TeV, the relic density at $w=10$ TeV is the same
as that at $w= 30$ TeV.

Similarly we investigate  contour plots for $0.1064 <
\Omega h^2 < 0.1176 $ (WMAP constraints) in the $M_U$ - $w$ plane.
The allowed region of $M_U$ is in the narrow band for fixing
$M_D$. For an
example, if $15< M_D < 21$ TeV, the allowed region of $M_U$ is in
40 $< M_U < $ 64 TeV.
%
 Finally we study contour plots for $0.1064 < \Omega h^2
< 0.1176 $ (WMAP constraints) in the $\la_1$ - $M_U$ plane.
Combining with the constraint on $\la_1$ given in (\ref{hu1}), we
can see that  the allowed bands are: $0.028< \la_1< 0.051$ for
$M_{D_2}$ = $M_{D_3}$=100 TeV, and  for $M_{D_2}$ =
$M_{D_3}$=12 TeV, $0.039< \la_1< 0.051$  if $w$ = 10 TeV, and no
region of $\la_1$ allowed if $w$ = 30 TeV.

 In next section we will study how to search for the DM in
the WMAP - allowed region, in direct and indirect searches.
We would like to analyze the results
as functions of $M_{H_1^0}$, which is expressed in term of
$\la_1$ and $w$. With the WMAP constrains, we use the best parameter space,
$\la_1$ = 0.04, $\la_2$ = 0.12,
 $\la_3 = -10^{-24}$, $\la_4$ = 0.06, $M_U$ = 36 TeV,
 $M_{D_2}$ = $M_{D_3}$ = 100 TeV,  and we vary 5 $< w <$ 15.3 TeV,
which requires $M_{H_1^0}$ to be few TeVs.

\section{Direct and indirect searches for the dark matter}
\label{sec5}

Experimentalists worldwide are actively chasing searches for
DM candidates either directly through detection of elastic
scattering of the weakly interacting massive particles with the nuclei in a large detector or
indirectly through detection of products of the dark matter annihilation
(photons, positrons, neutrinos or antiprotons) in the galaxy or
in the sun.

\subsection{Direct search}

In direct search, the recoil energy deposited by
scattering of WIMPs with the nuclei is measured. In general, WIMP-nuclei
interactions can be split  into two types: a spin independent interaction and a
spin dependent interaction. In our model, scalar $H_1^0$ Higgs DM
can only contribute to spin independent interaction.
To calculate the direct detection rate we use the method,
as mentioned in \cite{micrOmegas2.2};
The direct detection rate depends on the WIMP nucleus cross
section. To derive the $H_1^0$-nucleus cross section one has
to  compute first the interactions at the quark level then
convert them into effective
couplings of WIMPs to protons and neutrons. Finally, we have to
sum the proton and neutron contributions and turn this summation
into a cross section at the nuclear level.

The calculation of the
cross section for WIMP scattering on a nucleon is obtained at
the tree level. The normalized cross section on a
point-like nucleus is given as
\be \sigma^{SI}_{H^0_1N}=\frac{4\mu^2_{H^0_1}}{\pi}
\frac{(Zf_p+(A-Z)f_n)^2}{A^2}, \ee
where $\mu_{H^0_1}$ is the $H_1^0$ - nucleus reduced mass,
$f_p$ and $f_n$ are
amplitudes for protons and neutrons, respectively. For Xenon, $A$
= 131, $Z$ = 54, while for Germanium $A$ = 76, $Z$ = 32.
 The recoil spectrum of the nuclei depends on the velocity
distribution and is contained in the elastic form factor of the
nucleus. Using micrOMEGAs 2.4, we get the amplitudes and cross
sections for WIMP-nucleon elastic scattering calculated at zero
momentum as well as the total number of events/day/kg if we
consider detector made of Xe or Ge.

\begin{figure}[h]
\begin{center}
\includegraphics[width=7.5cm,height=6.5cm]{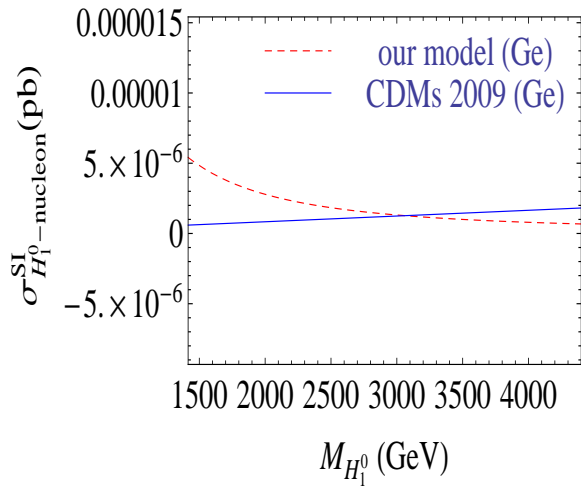}
\caption{\label{sigma}{$H_1^0$ DM-nucleon cross sections
 vs $M_{H_1^0}$ for $\la_1$ = 0.04, $\la_2$ = 0.12,
 $\la_3 = -10^{-24}$, $\la_4$ = 0.06, $M_U$ = 36 TeV and
 $M_{D_2}$ = $M_{D_3}$ = 100 TeV.}}
\end{center}
\end{figure}

 Fig. \ref{sigma} shows the values of $\sigma_{\rm WIMP-nucleon}$
as a function of dark matter mass by fixing the nucleon form
factors, $\sigma_0 = 40$ MeV and $\sigma_{\pi N}=55$ MeV. The value  of
$\sigma_{\rm WIMP-nucleon}$ is
in order of $10^{-6} (pb)$, which is allowed by experimental
constraints of CDMs 2009 (Ge). However, in the limit the dark matter
mass is smaller than 2.5 TeV or larger than 3.5 TeV, the
result given by our theoretical prediction is somehow different
from experiment of CDMs 2009 (Ge). We would like to emphasize that
the form factors of nucleons can be reset by changing the
pion-nucleon sigma term, $\sigma_{\pi N}$ = 55-73 MeV and from the
SU(3) symmetry breaking effect, $\sigma_0 = 35 \pm 5$ MeV
\cite{form}, however, the final WIMP-nucleon cross section predicted by our
model does not change much.
\begin{figure}[h]
\begin{center}
\includegraphics[width=5.5cm,height=6cm]{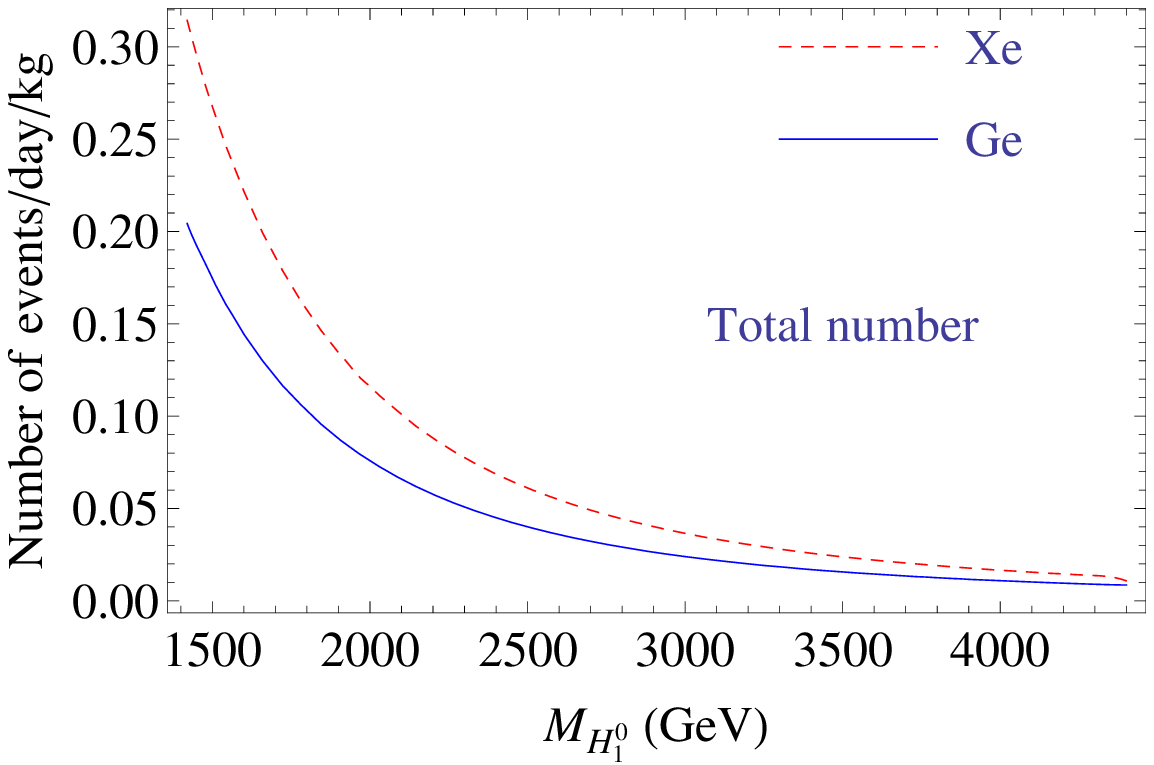}
\includegraphics[width=5.5cm,height=6cm]{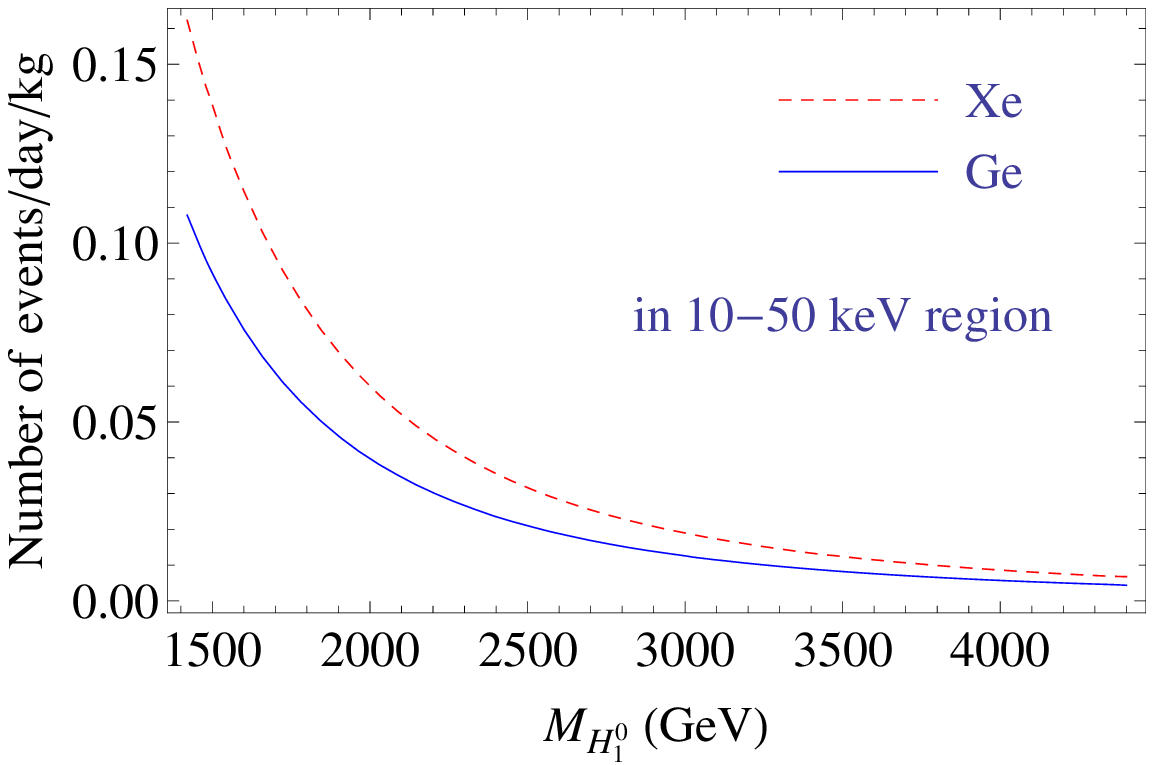}
\caption{\label{event}{Number of events/day/kg
 vs $M_{H_1^0}$ for $\la_1$ = 0.04, $\la_2$ = 0.12,
 $\la_3 = -10^{-24}$, $\la_4$ = 0.06, $M_U$ = 36 TeV and
 $M_{D_2}$ = $M_{D_3}$ = 100 TeV.}}
\end{center}
\end{figure}

Let us deal with the number of the nucleon recoil in Ge and Xe
detectors. The predicted number of nucleon recoil is given in
Fig. \ref{event}. The Xe detector is more sensitive than Ge
detector. In the limit, 2.5 $< M_{H_1^0} <$ 3.5 TeV, the
theoretical predictions are 22.3 and 14.6  nucleon recoils those
are observed in the Xe and Ge detectors for 1 kg per year,
respectively. It is worth mentioning that the number of the
nucleon recoils exposure between 10 keV $\sim$ 50 keV approximately
equals one half of total of that number.

\subsection{Indirect search}

\begin{figure}[h]
\begin{center}
\includegraphics[width=6.5cm,height=5.5cm]{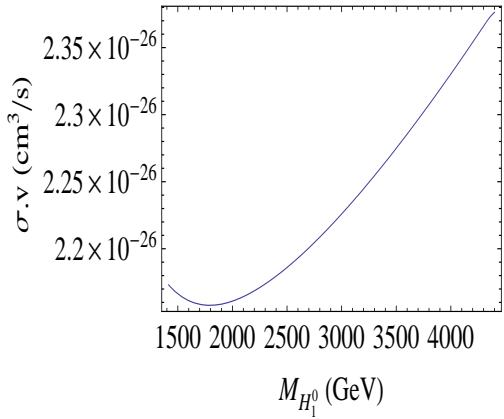}
\caption{\label{sigmaV}{ The annihilation cross section
times the relative velocity of incoming DM particles
 vs $M_{H_1^0}$ for $\la_1$ = 0.04, $\la_2$ = 0.12,
 $\la_3 = -10^{-24}$, $\la_4$ = 0.06, $M_U$ = 36 TeV and
 $M_{D_2}$ = $M_{D_3}$ = 100 TeV.}}
\end{center}
\end{figure}

DM annihilation in the galactic halo produces pairs of the SM
particles that hadronize and decay into stable particles. These
particles then evolve freely in the interstellar medium. The final
states with $\gamma, e^+$ and $\bar p$ are particularly
interesting as they are the subject of indirect searches. From
Feynman diagrams, we can see that the annihilation of $H_1^0H_1^0$
into $t \bar t$, $c \bar c$, $d \bar d$, $l\bar l$, $\nu_l
\bar{\nu_l}$ and $ZZ$ are done through s-channel $H^0$ exchange.
The annihilation of $H_1^0H_1^0$ into $H^0H^0$ is done through s-,
t-channel $H_1^0$, $H^0$ exchange or quartic couplings. Since the
couplings $H_1^0H_1^0H^0$,  $H_1^0H^0H^0$, and $H_1^0H_1^0H^0H^0$
are proportional to $\la_3$, the contributions of those
channels  are very small. With the choice of parameters
 $\la_1$ = 0.04, $\la_2$ = 0.12,
 $\la_3 = -10^{-24}$, $\la_4$ = 0.06, $M_U$ = 36 TeV,
 $M_{D_2}$ = $M_{D_3}$ = 100 TeV  and  $5 < w < 15.3 $ TeV,
the dominant channel is $ u \bar u$. For example, in case of
 $w= 10$ TeV, the relative
contribution in \% are displayed as following: ($97.40\% :$
$H_1^0H_1^0 \rightarrow u \bar u$); ($1.28 \% :$  $H_1^0H_1^0
\rightarrow b\bar b$); ($1.26 \% :$ $H_1^0H_1^0 \rightarrow s \bar
s$); ($0.05 \% :$ $H_1^0H_1^0 \rightarrow H_2^+ \bar H_2^-$); and
 ($0.01 \% :$ the rest).
The total annihilation cross section times the relative velocity
of incoming dark matter particles is shown in Fig. \ref{sigmaV}.
With the allowed region of the dark matter mass satisfied the WMAP
constraints, we find $2.15 \times 10^{-26} < \sigma v < 2.4 \times
10^{-26} cm^3/s$. This result is in the same order of that given
in \cite{sigmaV2}, which is said that away from the Higgs
resonance and the $W$ threshold, $<\sigma v>$ is essentially
constant and equal to the so-called typical annihilation cross
section, $<\sigma v> \sim 3.10^{-26} cm^3/s$. The AMS-2 experiment
given in \cite{sigmaV1} predicted for dark matter mass up to $600$
GeV. With dark matter mass from 100 GeV to 600 GeV, $\sigma v$
keeps constant value in order of $10^{-26} cm^3/s$. We expect that
our result for heavy dark matter can be covered by the future
experiment of   AMS-2.

 \begin{figure}[h]
 \begin{center}
 \includegraphics[width=5.5cm,height=5.5cm]{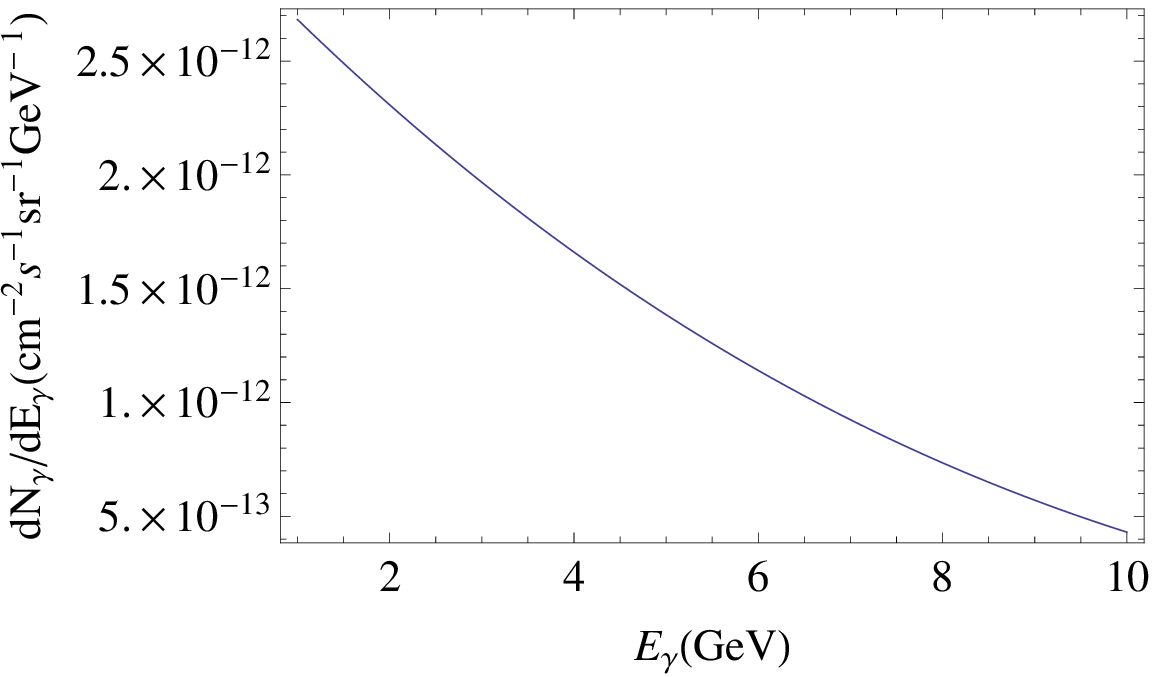}
  \includegraphics[width=5.5cm,height=5.5cm]{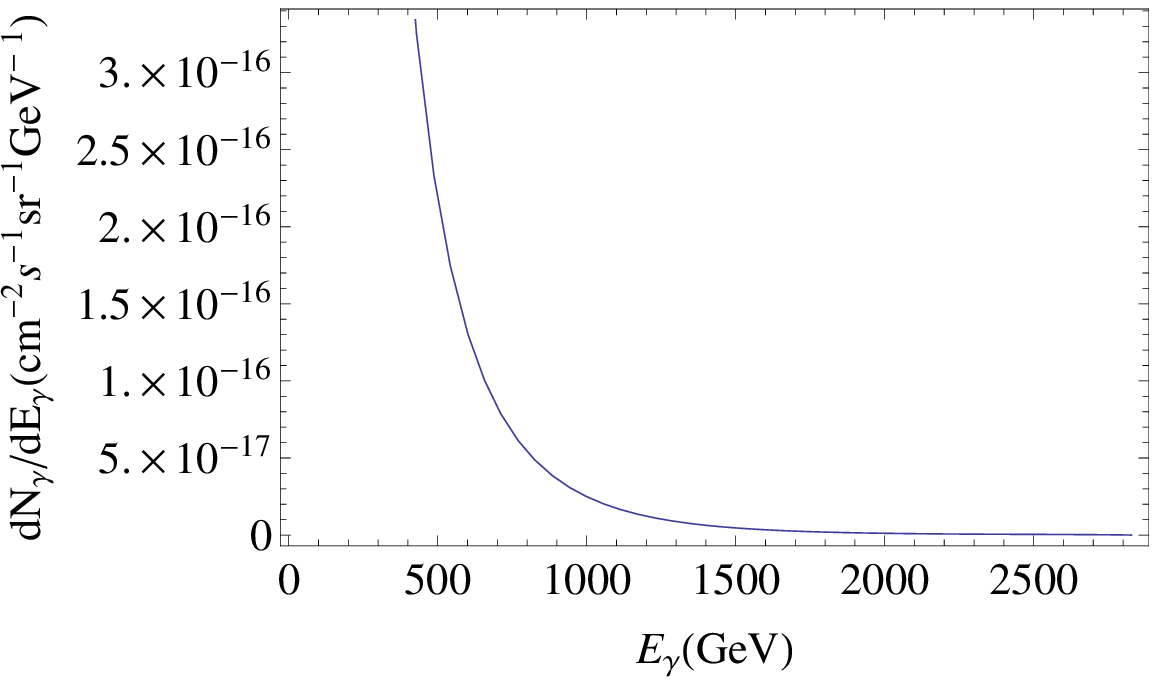}
 \caption{\label{photonFlux}{Photon flux vs $E$  for
$\la_1$ = 0.04, $\la_2$ = 0.12,
 $\la_3 = -10^{-24}$, $\la_4$ = 0.06,  $w$ = 10 TeV, $M_U$ = 36 TeV and
  $M_{D_2}$ = $M_{D_3}$ = 100 TeV.}}
 \end{center}
 \end{figure}

 Let us proceed with the discussion of the photon flux and energy
spectrum for dark matter with mass 2828.4 GeV. The spectrum for
photon flux is predicted in Fig. \ref{photonFlux}. It
is easy to see that the photon flux in the energy range from a few
MeV to $10$ GeV is much larger than that of higher energy
ranges. These results can be understood as following: As previously
mentioned, our model predicts that the annihilation of the
dark matter and anti-dark matter to $u$-quark and
$\overline{u}$-quark is the dominating channel. Therefore, the dominating jet is the
neutral pion jet, composed of pairs of $u$-quark and anti-quark in
this case. The $\gamma$-rays from particle annihilation processes
have spectra bounded by the rest mass energy of the annihilation
particle. The $\gamma$-rays are dominated by pion decay at low
energy from a few MeV to $10$ GeV. The additional contribution to
photon spectrum at higher energy due to other annihilation
processes such as polarization of the gauge bosons final state,
photon radiation, etc, which are predicted to be
tiny. Antiproton flux and positron flux also can be calculated by
MicrOMEGAs 2.4. The fluxes go down fast as functions of energy and
their values are significant at low energy as the same as photon
case.

\section{Conclusions}
\label{sec6}

We have shown that the economical 3-3-1 model provides a candidate
for dark matter without any discrete symmetry; and it just
requires some constraints on Higgs coupling constant. The scalar
Higgs $H_1^0$ is a good candidate for self interacting dark
matter. To forbid the decay of $H_1^0$, we require that $\la_1
\leq 0.051$,  $\la_1 < \la_4 $, and $|\la_3|\sim 10^{-24}$. The
constraint on the Higgs coupling $\la_3$ looks unnatural, which
could be canceled by introducing a discrete symmetry $S_3$. However,
by introducing new symmetry the Higgs sector becomes more complicated. Therefore, we do not
consider such scheme in this work. The parameter space has been studied in
detail and the results satisfying the WMAP observation
are summarized as following:
\begin{itemize}
  \item The relic density does not change much when varying
   $\la_2$ from 0.053 to 0.212, $\la_3$ around $-10^{-24}$ value,
   $\la_4$ from 0.004 to 200.
  \item $\la_1$ should be around 0.04.
  \item The region of $w$ is narrow
  compared to those of $M_U$, $M_{D_2}$ and  $M_{D_3}$.
  \item $U$-quark mass can be smaller or larger than $D$-quarks
  masses.
\end{itemize}
We have studied direct and indirect searches for $H_1^0$ dark
matter. The dark matter--nucleon cross section is in agreement
with CDMs 2009 (Ge). The total number of events observed in Xe and
Ge detectors is quite small because of the heavy dark
matter. We hope that these results can be covered in future by
experiments. Dark matter annihilation is considered with special
choice of parameters. In case $M_U< M_{D_2}= M_{D_3} $, the
dominant channel is $u\bar u$, while the dominant channel is
$s\bar s$ for
 $M_{D_2}= M_{D_3}>M_U $, because  the interactions of
quarks with other particles depend much on exotic quarks masses.
However, choosing  $U$- quark mass smaller or larger than
$D$- quark mass does not affect our results on cross section
times relative velocity as well as photon flux. The value of $\sigma v$
is in order of $10^{-26} cm^3/s$ in agreement with typical annihilation
cross section. Photon flux is dominated at low energy below 10 GeV.
\\

\section*{Acknowledgments}

N. T. T.  would like to thank S. Kraml at LPSC, Grenoble, France
for kind help and hospitality during her visit where this work was
initialed and P. V. Dong at Institute of Physics, Hanoi, Vietnam
for comments and H. Sung Cheon at Yonsei University, Seoul, Korea
for discussion. The work H. N. L. was supported in part by the
National Foundation for Science and Technology Development
(NAFOSTED) of Vietnam under Grant No. 103.01-2011.63. The work of
C. S. K. and N. T. T. was supported in part by the National
Research Foundation of Korea (NRF) grant funded by the Korea
government of the Ministry of Education, Science and Technology
(MEST) (No. 2011-0027275),  (No. 2011-0017430) and (No. 2011
-0020333).
\\


\newpage
\appendix
\section{APPENDIX: Feynman diagrams contributing to the annihilation of
$H_1^0$ dark matter} \label{appendix1}

\begin{figure}[htbp]
\begin{center}
\hspace*{-3cm}
\vspace*{-2cm}
\includegraphics[width=20cm,height=22cm]{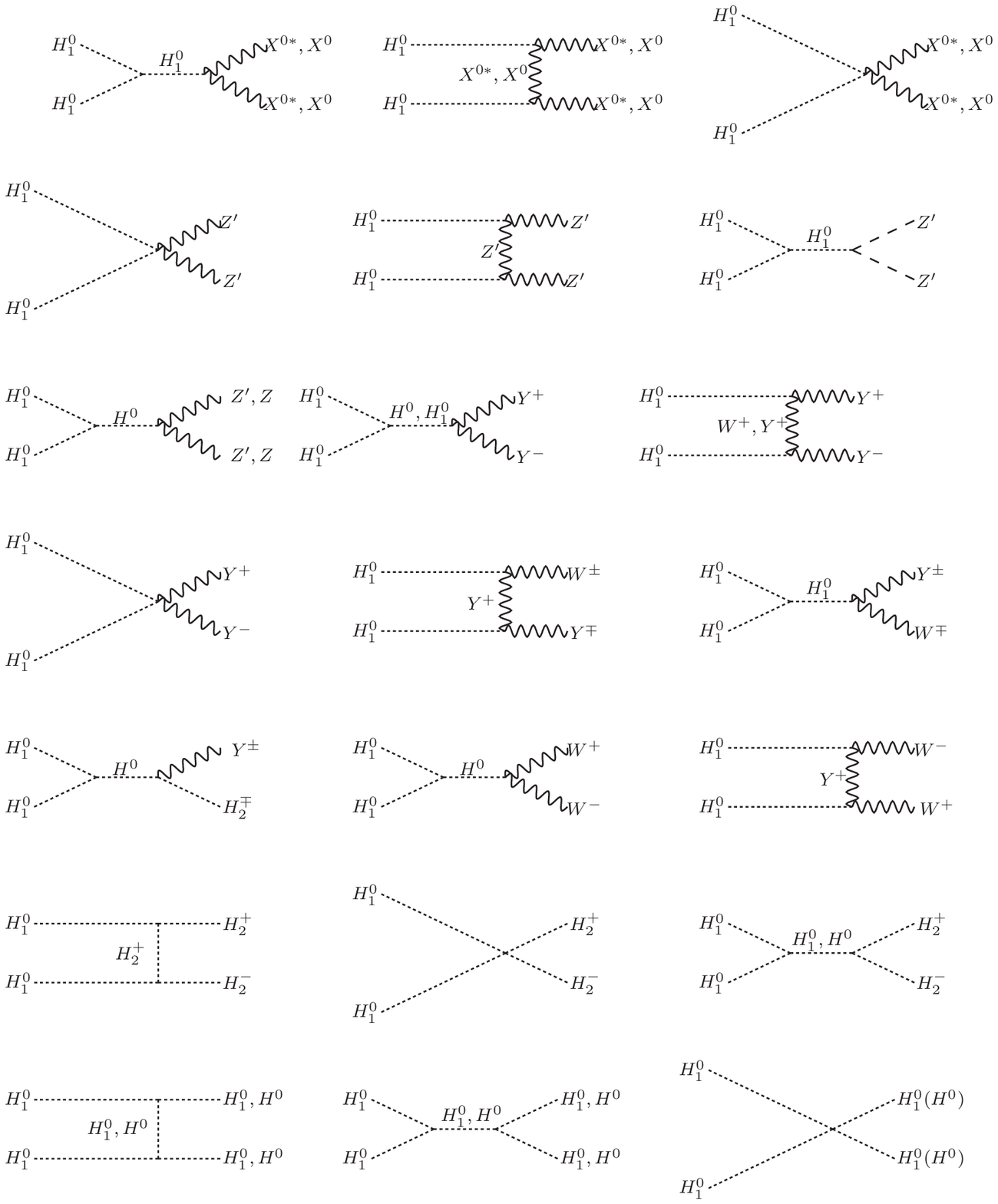}
\end{center}
\end{figure}

\newpage

\begin{figure}[htbp]
\begin{center}
\hspace*{-3cm}
\includegraphics[width=20cm,height=22cm]{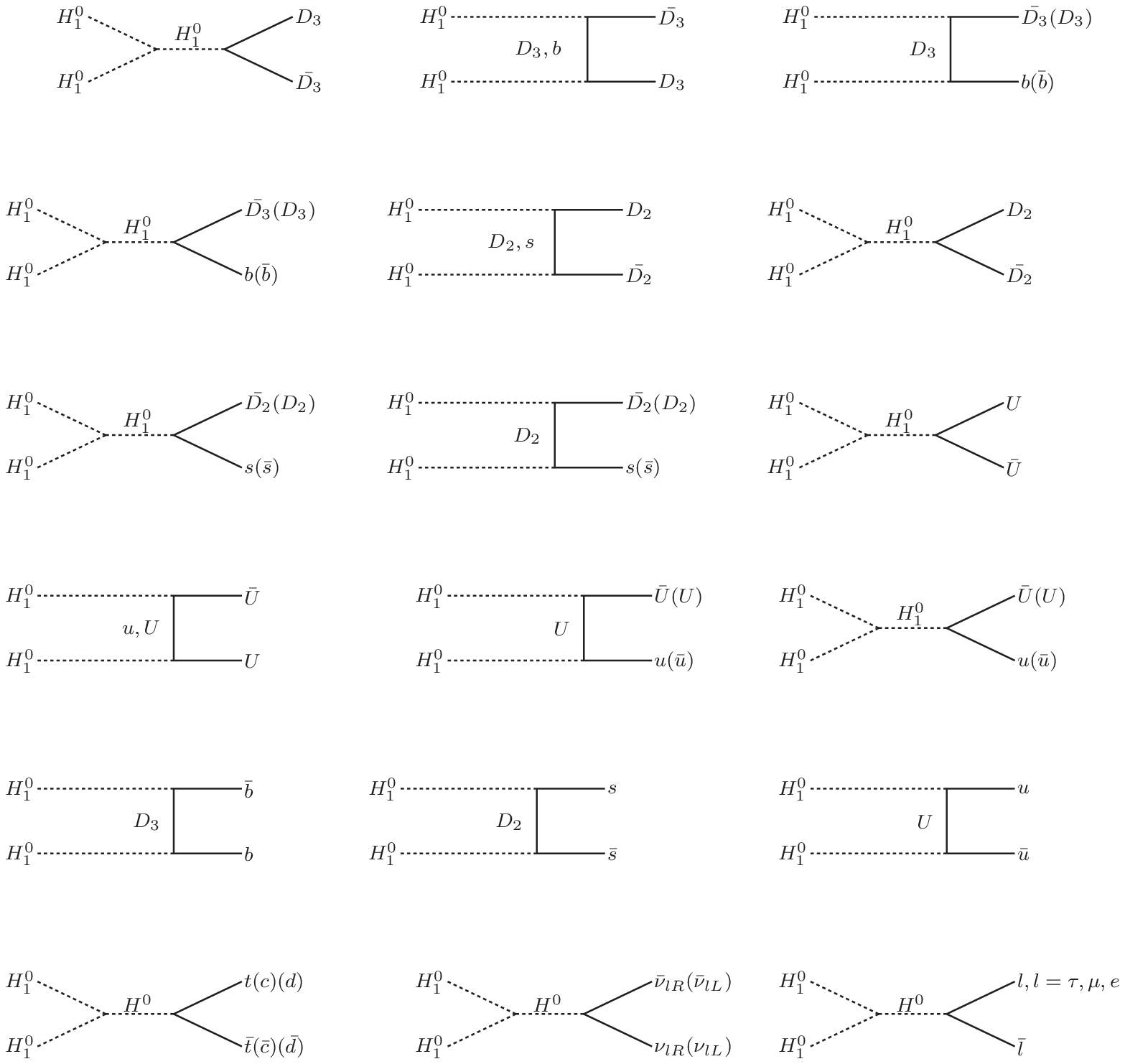}
\end{center}
\end{figure}

\end{document}